\documentclass[
journal=inoraj,
manuscript=article,
layout=twocolumn
]{achemso}

\usepackage[T1]{fontenc} 
\usepackage{bm}
\usepackage{siunitx}
\usepackage[version=4]{mhchem}
\usepackage{here}
\usepackage{comment}
\usepackage{autobreak}
\usepackage{amsmath}

\newcommand{\tit}[1]{\textit{#1}}
\newcommand{\trm}[1]{\textrm{#1}}
\newcommand{\mrm}[1]{\mathrm{#1}}

\author{Hiroki~Ninomiya}
\email{hiroki.ninomiya@aist.go.jp}
\phone{+81 (0)29 8612312}
\affiliation[AIST]
{National Institute of Advanced Industrial Science and Technology (AIST), Tsukuba, Ibaraki 305-8568, Japan}

\author{Terunari~Koshinuma}
\affiliation[AIST]
{National Institute of Advanced Industrial Science and Technology (AIST), Tsukuba, Ibaraki 305-8568, Japan}
\alsoaffiliation[TUS]
{Department of Physics, Graduate School of Science, Tokyo University of Science, Shinjuku, Tokyo 162-8601, Japan}
\author{Taichiro~Nishio}
\alsoaffiliation[TUS]
{Department of Physics, Graduate School of Science, Tokyo University of Science, Shinjuku, Tokyo 162-8601, Japan}

\author{Hiroshi~Fujihisa}
\affiliation[AIST]
{National Institute of Advanced Industrial Science and Technology (AIST), Tsukuba, Ibaraki 305-8568, Japan}

\author{Kenji~Kawashima}
\affiliation[AIST]
{National Institute of Advanced Industrial Science and Technology (AIST), Tsukuba, Ibaraki 305-8568, Japan}
\alsoaffiliation[IMRA]
{IMRA Material R\&D Co., Ltd., Kariya, Aichi 448-0032, Japan}

\author{Izumi~Hase}
\author{Shigeyuki~Ishida}
\author{Hiraku~Ogino}
\author{Akira~Iyo}
\author{Yoshiyuki~Yoshida}
\author{Yoshito~Gotoh}
\author{Hiroshi~Eisaki} 
\affiliation[AIST]
{National Institute of Advanced Industrial Science and Technology (AIST), Tsukuba, Ibaraki 305-8568, Japan}

\title{Experimental and Computational Determination of Optimal Boron Content in Layered Superconductor Sc$_{20}$C$_{8-x}$B$_x$C$_{20}$}

\begin{document}
\sloppy

\begin{abstract}
It is generally difficult to quantify the amounts of light elements in materials because of their low X-ray-scattering power, as this means that they cannot be easily estimated via X-ray analyses.
Meanwhile, the recently reported layered superconductor, Sc$_{20}$C$_{8-x}$B$_x$C$_{20}$, requires a small amount of boron, which is a light element, for its structural stability. 
In this context, here, we quantitatively evaluate the optimal $x$ value using both the experimental and computational approaches.
Using the high-pressure synthesis approach that can maintain the starting composition even after sintering, we obtain the Sc$_{20}$(C,B)$_{8}$C$_{20}$ phase by the reaction of the previously reported Sc$_{15}$C$_{19}$ and B (Sc$_{15}$B$_y$C$_{19}$).
Our experiments demonstrate that an increase in $y$ values promotes the phase formation of the Sc$_{20}$(C,B)$_{8}$C$_{20}$ structure; however, there appears to be an upper limit to the nominal $y$ value to form this phase.
The maximum $T_\mathrm{c}$ $(=7.6\text{\,K})$ is found to correspond with the actual $x$ value of $x \sim 5$ under the assumption that the sample with the same $T_\mathrm{c}$ as the reported value $(=7.7\text{\,K})$ possesses the optimal $x$ amount.
Moreover, we construct the energy convex hull diagram by calculating the formation enthalpy based on first principles. 
Our computational results indicate that the composition of Sc$_{20}$C$_4$B$_4$C$_{20}$ $(x=4)$ is the most thermodynamically stable, which is reasonably consistent with the experimentally obtained value.
\end{abstract}

\section{Introduction}
High-frequency lattice vibrations are known as one of the key factors that enhance the superconducting critical temperature $(T_\trm{c})$ within the framework of the Bardeen--Cooper--Schrieffer (BCS) theory\cite{BCS-theory,Ginzburg-BCS}. 
Prime examples reflecting the effects of lattice vibrations include the observation of the highest $T_\trm{c}$ in \ce{MgB2} at normal pressure\cite{MgB2-Nagamatsu,MgB2_BandStructure} and the discovery of nearly room temperature (RT) superconductivity in superhydride materials under exceedingly high pressures (HP)\cite{H2S_203K_nature,LaH10_250K_Nature,LaH10_260K_PRL}. 

\begin{figure}[!t]
\centering
\includegraphics[bb=0 0 554 348,width=8.6cm]{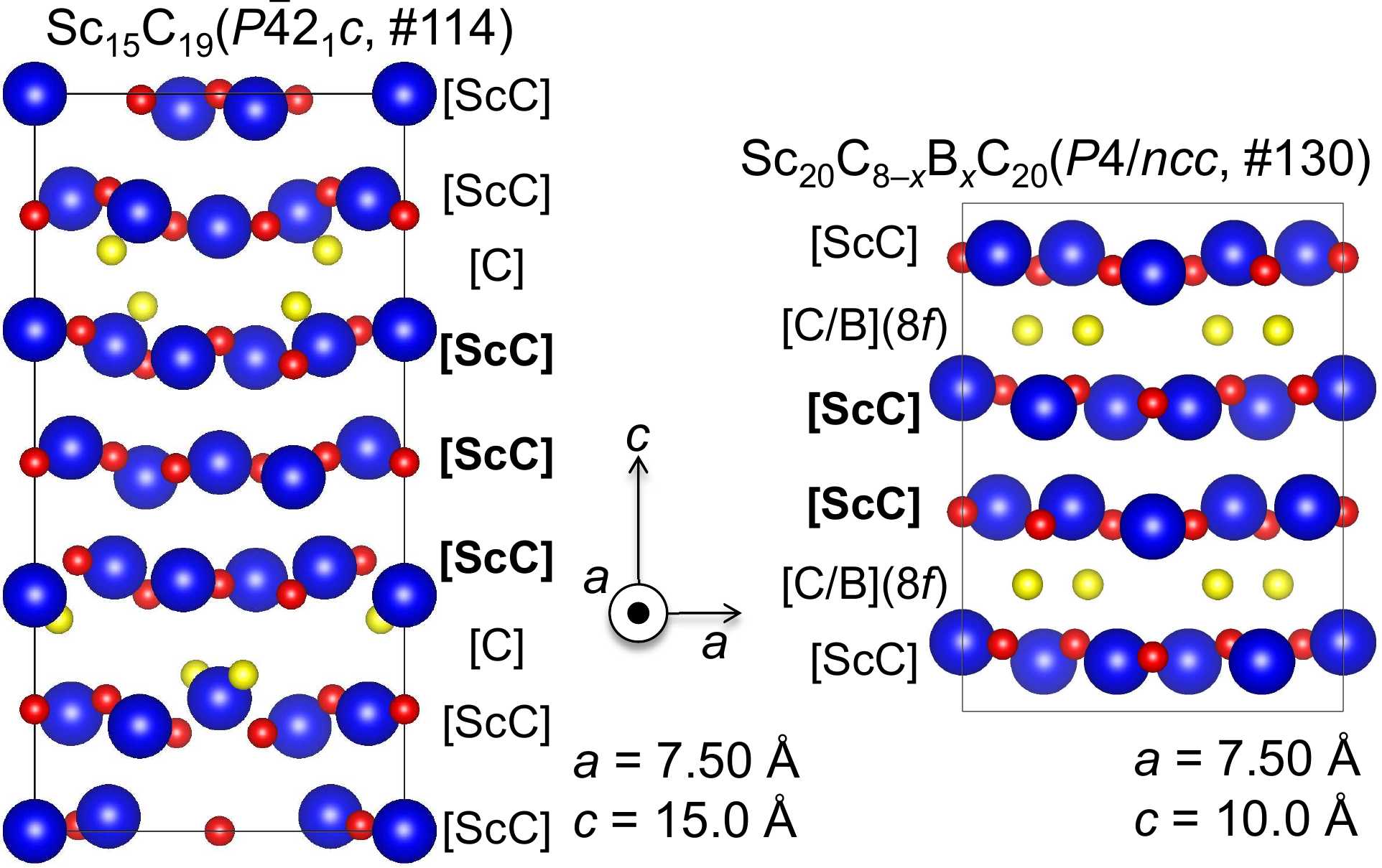}
\caption{
  Comparison of the crystal structures of (left) \ce{Sc15C19}\cite{Sc15C19} and (right) \ce{Sc20C_{8-x}B_{x}C20}\cite{Sc20C28-ninomiya} viewed from $a$-axis. 
  The larger and smaller spheres indicate \ce{Sc} and \ce{C} atoms, respectively. 
  The \ce{C} atoms in the \ce{C}-layer are shown in a different color (yellow) from those of the other sites.
  The structure was visualized using VESTA software\cite{VESTA}. 
  }
\label{cryst}
\end{figure}

Meanwhile, our search for superconducting materials containing light elements has recently yielded scandium borocarbide \ce{Sc20C_{8-x}B_{x}C20}, wherein the crystal structure is composed of alternately stacked \ce{ScC} and \ce{C/B} layers\cite{Sc20C28-ninomiya}. 
This compound exhibits bulk superconductivity at $T_\trm{c}=\SI{7.7}{K}$ with typical type-II behavior. 
In our studies, we found that a small amount of B $(x)$ is essential for phase stability although polycrystalline samples could be easily obtained by arc-melting each element.
In this regard, in a previous study, when B was excluded from the starting ingredients, the authors observed the formation of the non-superconducting \ce{Sc15C19} phase with a similar crystal structure\cite{Sc15C19}. 
Figure~\ref{cryst} compares the structures of the two materials. 
Both compounds exhibit a tetragonal layered structure with similar $a$-axis lengths and homogeneous atomic layers (ScC and C/B layers); however, there is a difference in their stacking sequence.
In \ce{Sc15C19} with the space group $P\bar{4}2_1c$, the C layer is separated by triple ScC layers; however, it can be observed that \ce{Sc15C19} exhibits an inexplicable buckling in each layer, particularly in the C layer. 
On the other hand, in \ce{Sc20(C,B)8C20} $(P4/ncc)$, double ScC layers are sandwiched between the C layers. 
Furthermore, using density functional theory (DFT) calculations for \ce{Sc20(C,B)8C20}, we previously estimated that the structural model, wherein B is substituted into the $8f$ site that forms the C layer, is the most stable configuration\cite{Sc20C28-ninomiya}. 

When an equivalent crystallographic site is occupied by different light elements, as in the formation of a solid-solution state, it is generally difficult to quantitatively analyze their concentrations.
This is because the lighter elements are relatively insensitive to direct-composition X-ray analysis techniques such as energy-dispersive X-ray spectroscopy (EDS) and electron probe microanalysis (EPMA). 
Indeed, an optimal amount of $x$ in \ce{Sc20C_{8-x}B_{x}C20} remains unknown because there is a slight discrepancy between the starting and actual compositions due to the evaporation of C and B during the melting and the formation of a secondary phase.
As reported in the literature\cite{Sc20C28-ninomiya}, considering the optimized starting composition, $x$ has been empirically assumed to be 1 or 2 at most; moreover, the $T_\mrm{c}$ value of the as-melted sample hardly depends on the nominal B content.
This implies that the arc-melting is unsuitable for precise composition tuning, which conversely suggests that the samples obtained by this method almost always have an unknown optimal content of B. 

In this study, we report the successful synthesis of \ce{Sc20(C,B)8C20} via a solid-state reaction under HP.
Although the reaction of the stoichiometric mixture of elemental Sc, B, and C powders yields no \ce{Sc20(C,B)8C20}, we demonstrate that this phase can be obtained by the reaction of \ce{Sc15C19} with the addition of B.
Because the starting composition is maintained during the HP synthetic process, this technique allows us to finely control the target composition.
Taking advantage of the benefits of HP synthesis in terms of high sealability of the sample, we attempt to estimate an optimal $x$ value in \ce{Sc20C_{8-x}B_{x}C20}.
Furthermore, to investigate the thermodynamic stability of the crystal structure as a function of $x$, we perform first-principles DFT calculations based on the convex hull approach\cite{ConvexHull-PRB2006}.

\section{Experimental and computational methods}
\subsection{Sample preparation}
Polycrystalline samples with a nominal composition of \ce{Sc15B_{y}C19}$(y=0-7.5)$ were
synthesized by means of the conventional solid-state reaction under pressures realized with the use of a cubic-anvil-type HP apparatus (CAP-07, Riken). 
The starting materials, powders of \ce{Sc15C19} and amorphous B, were weighed in a molar ratio of $1:y$, and then mixed using a zirconia mortar in a dry-nitrogen-filled glovebox. 
The precursor \ce{Sc15C19} was obtained by arc-melting Sc metal and graphite with a starting composition of $\ce{Sc}:\ce{C}=5:7$ on a water-cooled copper hearth.
The mixture of $\ce{Sc15C19}+\ce{B}$ was pressed into a pellet and encapsulated in a platinum crucible to avoid external contamination. 
The sample, assembled in an HP cell with pyrophyllite as the pressure-transmitting medium, was sintered at \SI{1300}{\degreeCelsius} for \SI{1}{h} under a pressure of \SI{4.3}{GPa} and subsequently quenched to room temperature $(\sim \SI{30}{\degreeCelsius})$.

\subsection{Characterization}
The phase purity of the obtained samples was confirmed by means of powder X-ray diffraction (XRD) with \ce{Cu}-$K_\alpha$ radiation at approximately \SI{293}{K}. 
The intensity data were collected by using a diffractometer (Ultima-IV, Rigaku) equipped with a high-speed X-ray detector (D/teX Ultra, Rigaku). 
For the refinement of the crystal structure of \ce{Sc15C19}, we performed the Rietveld analysis of the data of the as-melted sample using the BIOVIA Materials Studio Reflex software package (version 2018 R2)\cite{BIOVIA}. 
Single-phase analysis was adopted in this study. 

We examined the sample superconductivity and $T_\trm{c}$ by means of magnetization $(M)$ measurements using a commercial SQUID magnetometer (MPMS-XL, Quantum Design), with the temperature $(T)$ ranging from \SI{1.8}{K} to \SI{10}{K} under a magnetic field $(H)$ of \SI{10}{Oe}. 
The relevant data were acquired for both the zero-field cooling and field cooling processes. 

\subsection{Enthalpy calculations}
To construct the convex hull diagram, we performed structural optimization for each $x$ value of \ce{Sc20C_{8-x}B_{x}C20} $(x=0-8)$ using density functional theory (DFT) calculations and estimated the enthalpy of formation. 
We used the on-the-fly ultrasoft pseudopotentials\cite{OTF-pseudopotential_PRB} generated by the CASTEP code\cite{CASTEP} along with GGA-PBE exchange-correlation functionals\cite{GGA}. 
The cut-off energy for the plane-wave basis set was \SI{460}{eV} for these calculations. 
The $k$-space was sampled by a $3\times3\times2$ Monkhorst-Pack grid\cite{MP-grid}. 
In this study, we created structural models in which the C atoms at the $8f$ site are selectively replaced by B atoms, and therefore, we removed the constraints of the space group $P4/ncc$(\#130) and set it to $P1$(\#1).

\section{Results}
\subsection{Material synthesis and phase identification}
\subsubsection{Reexamination of crystal structure of \ce{Sc15C19}}
\begin{figure}[t]
\centering
\includegraphics[bb=0 0 501 423,width=8.6cm]{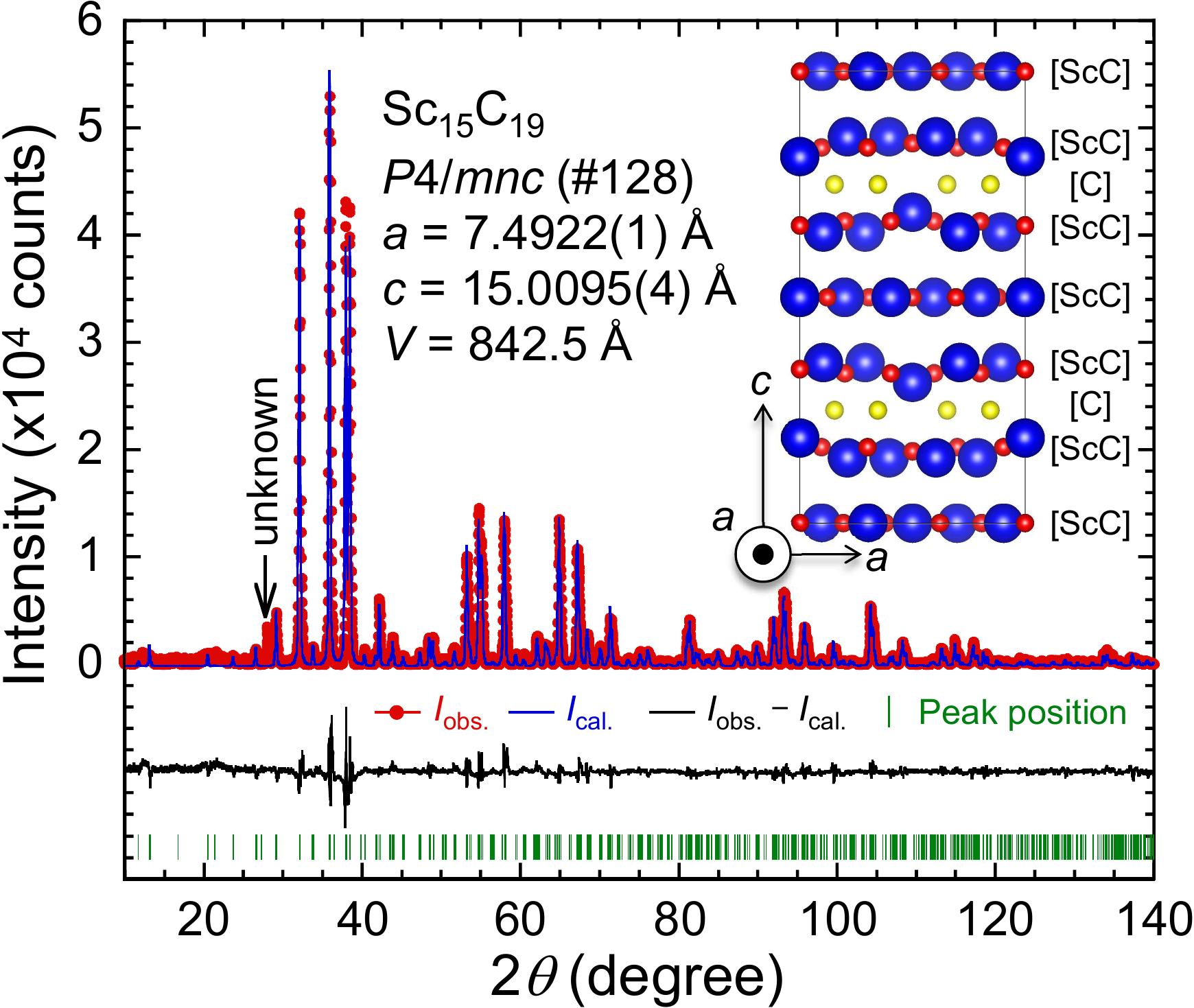}
\caption{
  Rietveld analysis results for as-melted \ce{Sc15C19}. 
  The observed and calculated intensity data are denoted by $I_\trm{obs.}$ and $I_\trm{cal.}$, respectively. 
  The difference curve (in black) $(I_\trm{obs.}-I_\trm{cal.})$ is shown with a downward shift of $1\times10^4$ counts.
  The typical peak originating from unknown phases is indicated by the arrow.
  The inset illustrates the re-refined structure with the space group $P4/mnc$.
  }
\label{Sc15C19-rietveld}
\end{figure}

In the next section, we show that the \ce{Sc20C_{8-x}B_{x}C20} phase was indeed obtained by the HP synthesis of the mixture of \ce{Sc15C19} and B powders. 
In this study, before the HP synthesis, we reexamined the X-ray structure of \ce{Sc15C19}. 
First, the structural optimization of the previous $P\bar{4}2_1c$ model was performed by using DFT calculations, which resulted in a large decrease in the formation enthalpy per unit cell from $\SI{-44203.0}{eV}$ to $\SI{-44228.8}{eV}$. 
From the inset of Fig.~\ref{Sc15C19-rietveld}, it can be observed that the degree of buckling in the C layer clearly reduces in the obtained model with the higher symmetry of $P4/mnc$ (\#128).
The enthalpy difference of about \SI{26}{eV} between two structural models might be due to the close proximity between the C atoms in the C and ScC layers in the originally reported structure, as can be seen from the left panel of Fig.~\ref{cryst}.
Figure~\ref{Sc15C19-rietveld} shows the result of the Rietveld analysis applied to \ce{Sc15C19}. 
The experimental XRD pattern is well-fitted by the $P4/mnc$ model, which yields a weighted-profile factor $(R_\trm{wp})$ and expected reliability factor $(R_\trm{e})$ of $R_\trm{wp}=11.45\%$ and $R_\trm{e}=10.04\%$, respectively.
These values correspond to a goodness-of-fit indicator $(S)$ of $S=1.14$, thereby confirming the appropriateness of the analysis.
The fractional coordinates are summarized in Table~\ref{15-19atompositon}. 
Additionally, as indicated by the arrow in Fig.~\ref{Sc15C19-rietveld}, a non-negligible peak originating from a certain secondary phase is observed at around $30^\circ$; however, we found that the origin of this reflection cannot be explained even by the originally reported $P\bar{4}2_1c$ model.

\begin{table*}
\caption{
Atomic coordinates and isotropic displacement parameters $(U_\trm{iso})$ for \ce{Sc15C19} refined by Rietveld analysis at room temperature. 
The site occupancy for each site was fixed at 1.
}
\label{15-19atompositon}
\begin{tabular}{ccllll} \hline
Site & \begin{tabular}[c]{@{}c@{}}Wyckoff\\ position\end{tabular} & \multicolumn{1}{c}{\textit{x}} & \multicolumn{1}{c}{\textit{y}} & \multicolumn{1}{c}{\textit{z}} & \multicolumn{1}{c}{$U_\trm{iso}$\textsuperscript{\emph{a}}} \\ \hline
Sc1 & $16i$ & 0.2108(4) & 0.3956(4) & 0.1447(2) & 0.013(1) \\
Sc2 & $4e$ & 0 & 0 & 0.1886(5) & 0.013 \\
Sc3 & $8h$ & 0.3020(6) & 0.0955(7) & 0 & 0.013 \\
Sc4 & $2b$ & 1/2 & 1/2 & 0 & 0.013 \\
C1 & $16i$ & 0.3005(16) & 0.0961(16) & 0.1684(5) & 0.013 \\
C2 & $4e$ & 1/2 & 1/2 & 0.1591(16) & 0.013 \\
C3 & $8h$ & 0.1931(21) & 0.3781(22) & 0 & 0.013 \\
C4 &$ 2a$ & 0 & 0 & 0 & 0.013 \\
C5 &$ 8g$ & 0.3459(20) & 0.1541(20) & 1/4 & 0.013 \\ \hline
\end{tabular}

\textsuperscript{\emph{a}} A global $U_\trm{iso}$ factor was employed for all crystallographic sites. 
\end{table*}

\subsubsection{High-pressure synthesis of \ce{Sc15B_{y}C19}}
\begin{figure}[t]
\centering
\includegraphics[bb=0 0 357 502,width=8.6cm]{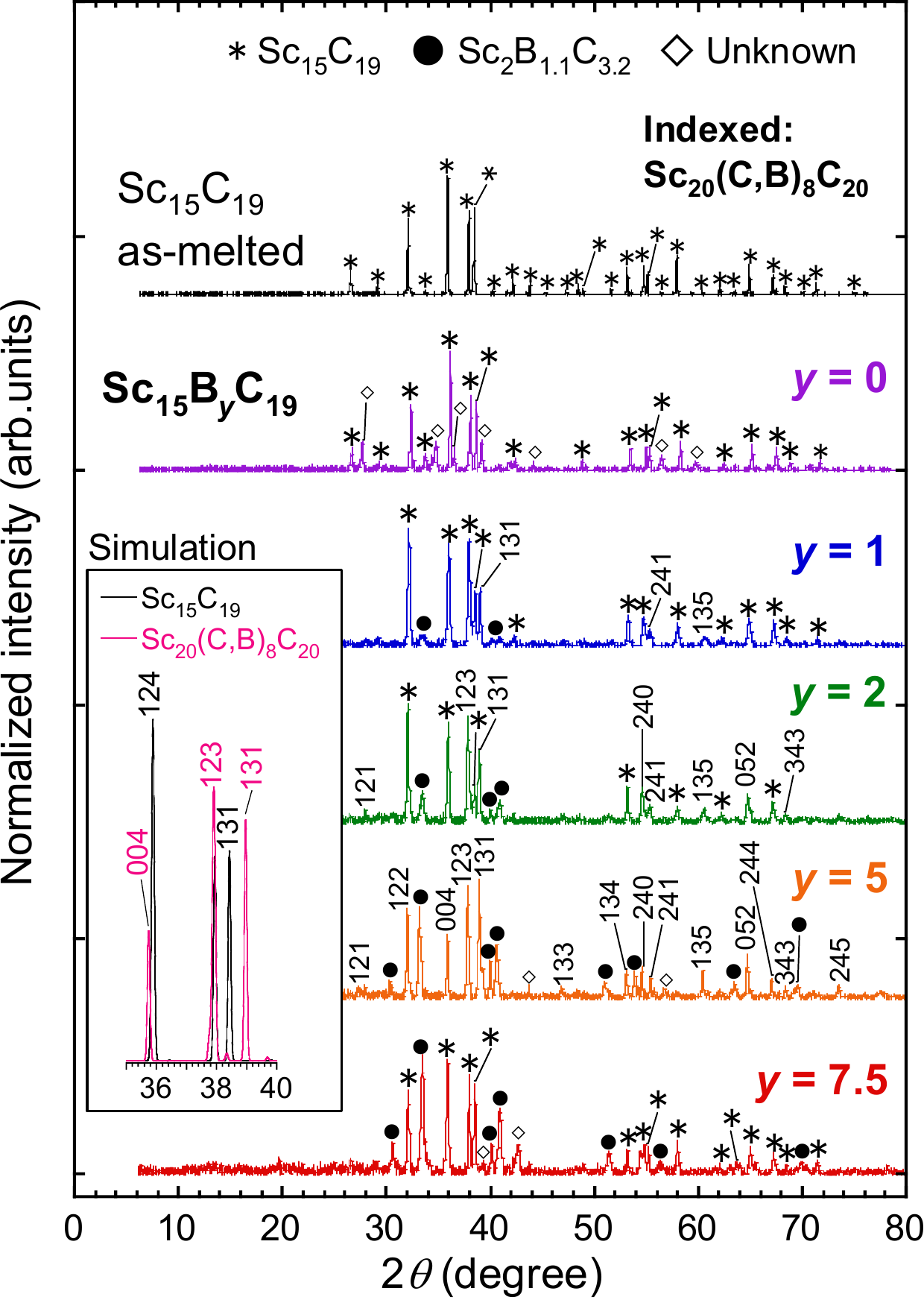}
\caption{
  Experimental X-ray diffraction (XRD) profiles of \ce{Sc15B_{y}C19}$(y=0,1,2,5,7.5)$ and the as-melted \ce{Sc15C19} before high pressure (HP) synthesis. 
  All patterns were normalized to their respective maximum intensities and offset for good visibility. 
  The inset shows the enlarged view near the main peak of the simulated patterns of the \ce{Sc15C19} and \ce{Sc20(C,B)8C20} structures.
  }
\label{xrd}
\end{figure}
Figure~\ref{xrd} depicts the XRD measurement results of \ce{Sc15B_{y}C19} $(y=0-7.5)$. 
For comparison, the data of the as-melted \ce{Sc15C19} before HP synthesis are also shown. 
The pattern of the sample sintered under HP without B, i.e., the $y=0$ specimen, is found to originate from the \ce{Sc15C19}-type structure, although a few unknown peaks are observed. 
The peak positions, which were labeled by the \ce{Sc15C19} phase in the $y=0$ sample, are uniformly shifted slightly to the wide-angle side relative to those of the as-melted compound, corresponding to a 1.4\% lattice shrinkage.
The addition of B to \ce{Sc15C19}, namely, increasing $y$, affords a gradual change in the XRD pattern. 
In the samples with $y>0$, the XRD peaks originating from the \ce{Sc20(C,B)8C20} structure are observed.
Because of the structural similarity between \ce{Sc15C19} and \ce{Sc20(C,B)8C20}, their XRD patterns are quite similar. 
To examine the distinctions between the two patterns, we present the enlarged view around the main peak in the inset of Fig.~\ref{xrd}.
The most significant difference is the position of the (131) peak, which is observed around $2\theta=38^\circ-39^\circ$. 
In comparison with the case of \ce{Sc15C19}, this peak from the \ce{Sc20(C,B)8C20} structure shifts to a wider angle. 
Furthermore, the maximum peak of \ce{Sc15C19} appears at $36.0^\circ$(124 peak), whereas that of \ce{Sc20(C,B)8C20} is observed at $38.0^\circ$(123 peak).
As $y$ increases, the peak intensity of \ce{Sc20(C,B)8C20} increases relative to that of \ce{Sc15C19}; furthermore, peaks derived from another non-superconducting impurity \ce{Sc2B_{1.1}C_{3.2}}\cite{Sc2B1.1C3.2} exhibit a gradual strengthening in intensity. 
At $y=7.5$, all of the peaks originating from the \ce{Sc20(C,B)8C20} structure suddenly disappear, and the XRD pattern can be assigned by \ce{Sc15C19}, \ce{Sc2B_{1.1}C_{3.2}}, and unknown phases. 

\begin{figure}[t]
\centering
\includegraphics[bb=0 0 558 477,width=8.6cm]{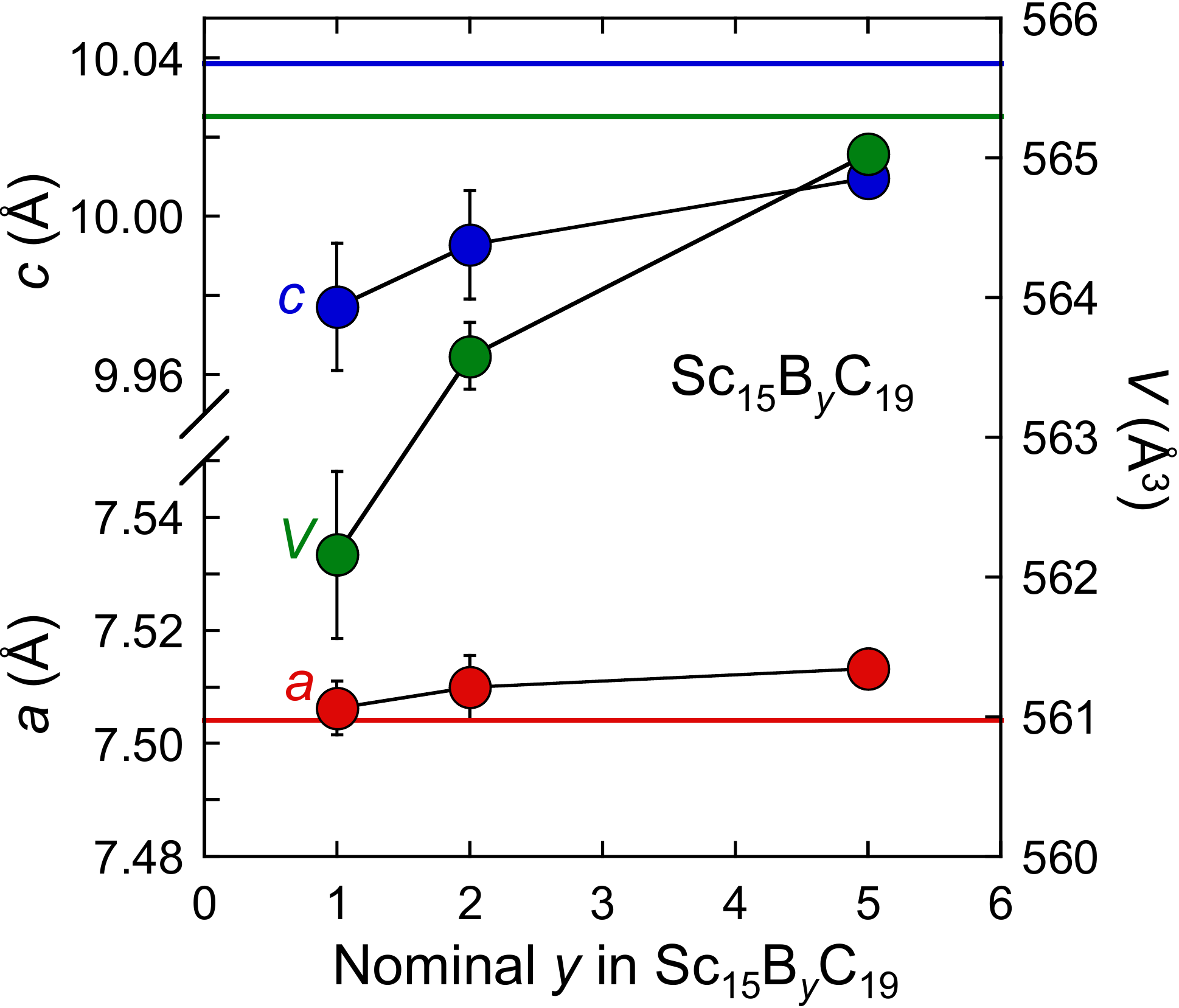}
\caption{
  Variations in lattice constants $(a, c)$ and cell volume $(V)$ with respect to nominal $y$ in \ce{Sc15B_{y}C19}, calculated by assuming the \ce{Sc20(C,B)8C20} structure.
  The standard deviation is denoted as an error bar.
  The horizontal color lines represent the corresponding values for \ce{Sc20(C,B)8C20} synthesized by arc-melting\cite{Sc20C28-ninomiya}. 
  }
\label{acV}
\end{figure}

Figure~\ref{acV} shows the evolution of the cell parameters of \ce{Sc15B_{y}C19} as functions of $y$. 
In the study, the lattice constants were calculated by means of the least-squares method by employing the peaks characterized by the \ce{Sc20(C,B)8C20} structure. 
As the nominal $y$ increases, the $a$- and $c$-axis lengths are extended slightly by $\sim 0.1\%$ and $\sim 0.3\%$, respectively, resulting in a $\sim 0.5\%$ expansion of the cell volume $V$. 
This indicates an increase in the B content in the \ce{Sc20(C,B)8C20} structure, and the expansion of $V$ can be attributed to the difference between the atomic radii of \ce{B} and \ce{C}.
The cell volume of the $y=5$ sample is comparable with that obtained by arc-melting\cite{Sc20C28-ninomiya}. 

\subsection{Superconductivity}
\begin{figure}[t]
\centering
\includegraphics[bb=0 0 495 390,width=8.6cm]{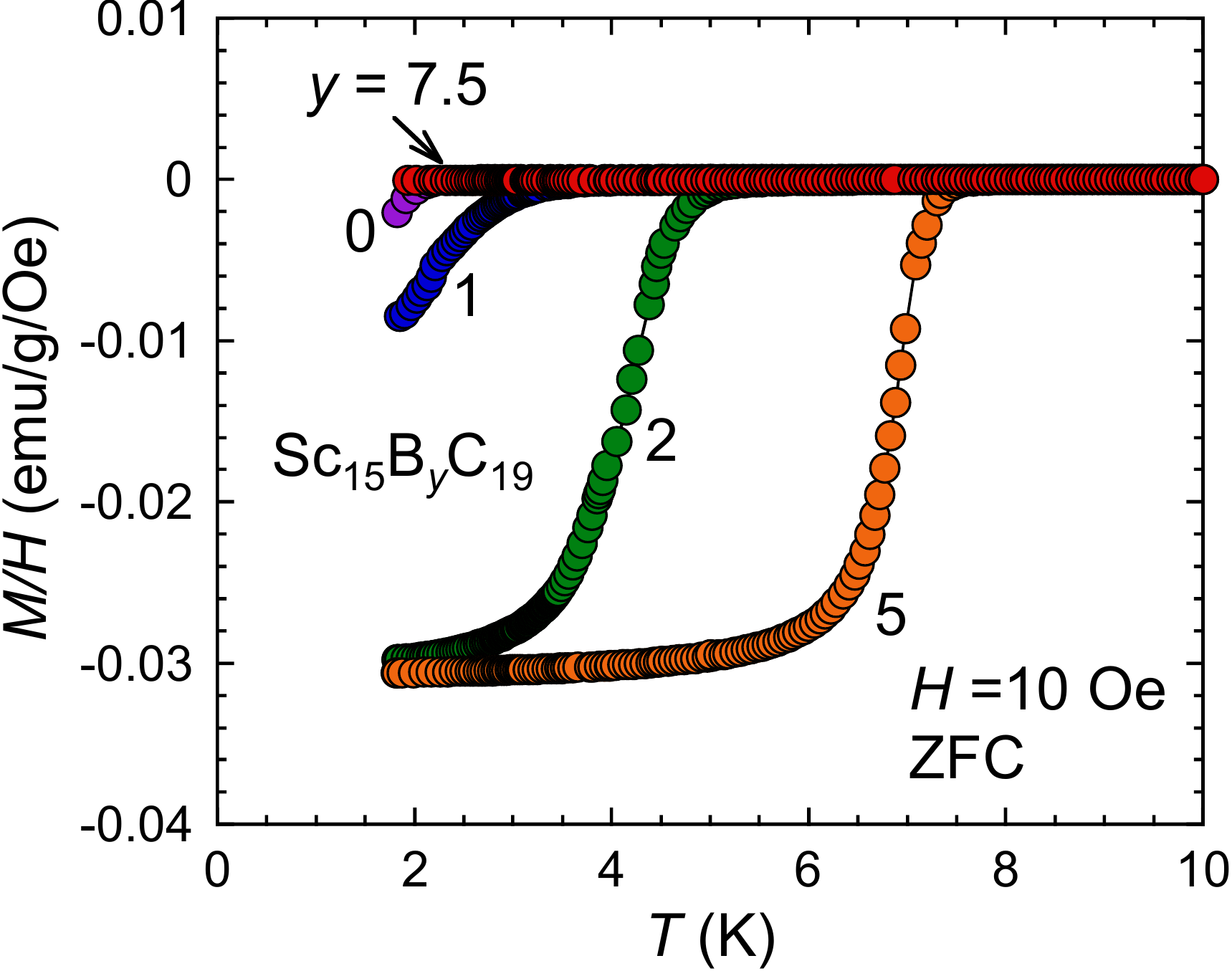}
\caption{
  Temperature dependence of magnetic susceptibility $(M/H)$ for samples with the nominal composition of \ce{Sc15B_{y}C19}$(y=0-7.5)$ measured under field of $H=\SI{10}{Oe}$ with the ZFC mode.
  }
\label{MT}
\end{figure}

The magnetic susceptibility $(M/H)$ vs. $T$ data are plotted in Fig.~\ref{MT}. 
A trace of superconductivity is observed at the onset of $T_\trm{c}=\SI{2.2}{K}$ even in the sample with $y=0$. 
This result suggests the formation of the \ce{Sc20(C,B)8C20} phase devoid of B; however, the origin of this superconductivity is currently unclear, because there is no reproducibility in the presence or absence of superconductivity of the B-free samples. 
With increase in $y$, $T_\trm{c}$ increases gradually, and the diamagnetic signal corresponding to the superconducting transition becomes sharper. 
The maximum $T_\trm{c}$ of \SI{7.6}{K} is observed for the $y=5$ specimen, and this value closely agrees with that for the arc-melted \ce{Sc20(C,B)8C20}\cite{Sc20C28-ninomiya}. 
For $y=7.5$, the $M/H$ curve shows no anomalies down to \SI{1.8}{K}.
Considering that no XRD peaks attributed to the \ce{Sc20(C,B)8C20} structure are observed, as illustrated in Fig.~\ref{xrd}, the superconductivity confirmed in the other samples does not originate from \ce{Sc15C19}, \ce{Sc2B_{1.1}C_{3.2}}, nor the unknown phases observed in the XRD data of the $y=7.5$ sample.

\subsection{Structural optimization and enthalpy evaluations}
\begin{figure}[t]
\centering
\includegraphics[bb=0 0 381 488,width=8.6cm]{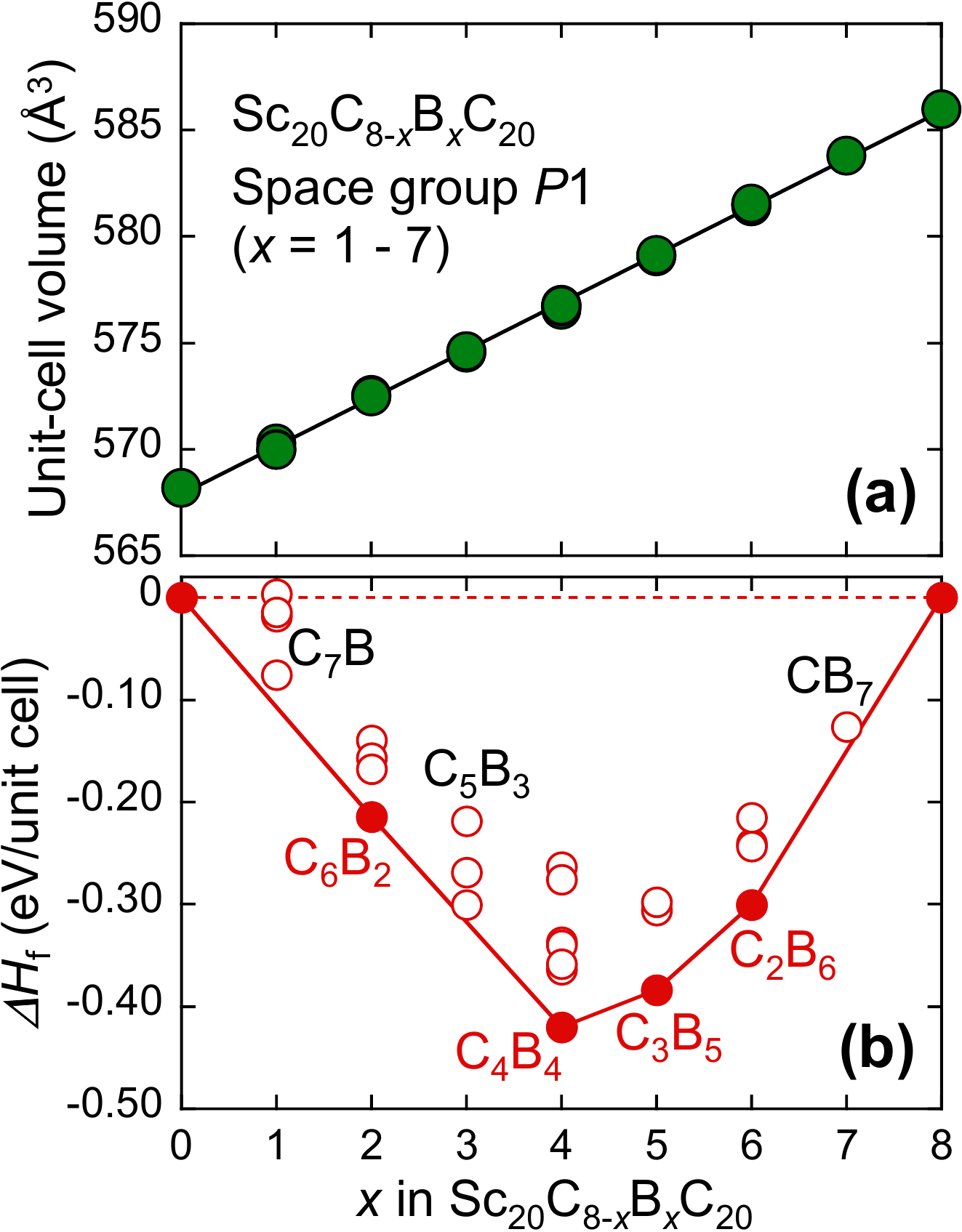}
\caption{
  (a) Optimized cell volume and (b) enthalpies of formation as functions of $x$ in \ce{Sc20C_{8-x}B_{x}C20}.
  For all $x$ values, the basic lattice structures of the crystal model used in the optimization are identical to each other.
  The several data points for each $x$ value indicate the different possible ways in which B atoms are substituted into each C layer. 
  The solid line in (b) denotes a convex hull on which the data are represented by closed circles, while the data above the convex hull are indicated by open circles. 
  }
\label{convex}
\end{figure}

To optimize the \ce{Sc20C_{8-x}B_{x}C20} structure for various $x$ values from the computational point of view, we performed \tit{ab initio} DFT calculations for $x=0-8$ in increments of one unit.
For $x=1-7$, the $P4/ncc$ symmetry was intentionally lowered to $P1$ without changing the basic lattice structure; we employed the structural model where, one by one, C atoms at the $8f$ site are replaced by B.
Figures~\ref{convex}(a) and (b) illustrate the resultant cell volume and the corresponding formation enthalpies per unit cell for each $x$ value, respectively.
The enthalpy data are plotted as the difference $(\varDelta H_\trm{f})$ on the basis of those for $x=0$.
Generally, such a plot is referred to as the energy convex hull diagram\cite{ConvexHull-FluidPhaseEquilib2017,ConvexHull-PRE2017}.
For example, this technique allows us to discuss the phase stability with respect to differences in the content between each in multi-component systems\cite{SnS-etc_ConvexHull2012,HEA_ConvexHull_PRX2015,PHx-ConvexHull-PRB2016}.
In this paper, we examined the most stable B composition in \ce{Sc20(C,B)8C20} assuming the B-C replacement in the C layer.
As plotted in Fig.~\ref{convex}(a), the cell volume expands linearly with increasing $x$, which is qualitatively consistent with the experimental behavior shown in Fig.~\ref{acV}.
Meanwhile, in Fig.~\ref{convex}(b), the compositions corresponding $x=2,4,5,$ and 6 on the convex hull are thermodynamically stable at zero temperature, which means that $x=1,3,$ and 7 affords a relatively metastable or unstable compound. 
Among these compounds, \ce{Sc20C4B4C20} $(x=4)$ with the lowest $\varDelta H_\trm{f}$ was considered to be the most stable in this system. 
Our model construction implies that upon viewing the structure from the $c$-axis, the formation enthalpy becomes higher when the B atoms in the C layer at $z=0.25$ and 0.75 show a tendency to overlap with each other.
This result is probably due to the local stress in the $c$-axis direction arising from the atomic size difference between B and C atoms. 

\section{Discussion}
\begin{figure}[t]
\centering
\includegraphics[bb=0 0 498 389,width=8.6cm]{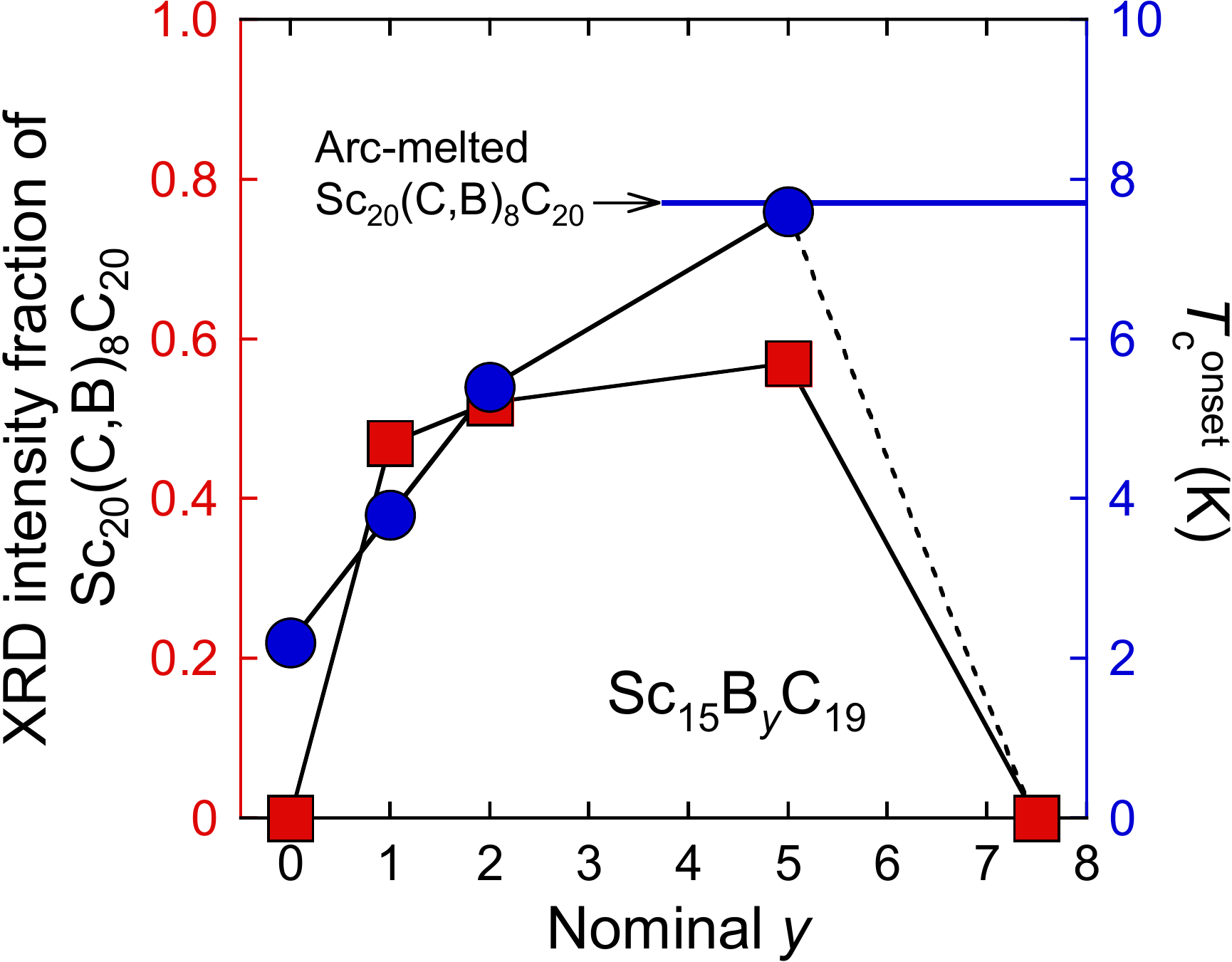}
\caption{
The $y$-dependent XRD intensity fraction of the (131) peak of the \ce{Sc20(C,B)8C20} structure (squares) and $T_\trm{c}^\trm{onset}$ (circles) of \ce{Sc15B_{y}C19} $(y=0-7.5)$.
The value on the left $y$-axis is defined as
\footnotesize
$I_{131}^{\ce{Sc20(C,B)8C20}}/\left(I_{131}^{\ce{Sc20(C,B)8C20}}+I_{131}^{\ce{Sc15C19}}+ I_\trm{max}^{\ce{Sc2B_{1.1}B_{3.2}}}\right)$.
\normalsize
For comparison, $T_\trm{c}$ of the arc-melted sample\cite{Sc20C28-ninomiya} is denoted by the horizontal line.
}
\label{y-Tc}
\end{figure}

We demonstrated that the \ce{Sc20C_{8-x}B_{x}C20} phase is obtained from $\ce{Sc15C19}+y\ce{B}$ with the precise control of $y$ using the HP synthesis approach.
Although an increase in the nominal $y$ value yielded the impurity \ce{Sc2B_{1.1}C_{3.2}}, the XRD peak intensity of \ce{Sc20(C,B)8C20} increased relative to that of \ce{Sc15C19}. 
Figure~\ref{y-Tc} depicts the XRD intensity fraction of \ce{Sc20(C,B)8C20} in the various \ce{Sc15B_{y}C19} samples, which was defined by calculating
\footnotesize
\begin{math}
I_{131}^{\ce{Sc20(C,B)8C20}}/\left(I_{131}^{\ce{Sc20(C,B)8C20}}+I_{131}^{\ce{Sc15C19}}+ I_\trm{max}^{\ce{Sc2B_{1.1}B_{3.2}}}\right),
\end{math}
\normalsize
where $I_{131}^{\ce{Sc20(C,B)8C20}}$ and $I_{131}^{\ce{Sc15C19}}$ represent the (131) peak intensities from \ce{Sc20(C,B)8C20} and \ce{Sc15C19}, respectively, and $I_\trm{max}^{\ce{Sc2B_{1.1}B_{3.2}}}$ corresponds to the maximum peak intensity of the \ce{Sc2B_{1.1}C_{3.2}} phase at around $2\theta = 33^\circ$ in Fig.~\ref{xrd}. 
The confirmed $T_\trm{c}^\trm{onset}$ is also plotted in the figure. 
Here, it is noteworthy that the nominal $y$ on the horizontal axis does not necessarily correspond to $x$ in \ce{Sc20C_{8-x}B_{x}C20} because the samples with $y>0$ contain the B-containing secondary phase originating from \ce{Sc2B_{1.1}C_{3.2}}. 
As $y$ increases, the fraction and $T_\trm{c}$ exhibit qualitatively similar trends. 
At $y=5$, at which the fraction reaches its maximal value, $T_\trm{c}$ also exhibits a maximum of \SI{7.6}{K}. 
No \ce{Sc20(C,B)8C20} phase is confirmed in the sample with $y=7.5$, which implies the existence of an upper $y$ limit value for the formation of this structure.

As mentioned in the introduction, we previously reported that the arc-melted sample may always contain an optimal B content\cite{Sc20C28-ninomiya}, because its $T_\trm{c}$ value $(=\SI{7.7}{K})$ does not depend significantly on the starting composition of B.
This means that it is reasonable to assume that a sample with similar cell volume and $T_\trm{c}$ to the arc-melted sample has the optimal $x$ value in \ce{Sc20C_{8-x}B_{x}C20}.
Among the obtained samples in this study, only the $y=5$ sample is found to satisfy this assumption, as depicted in Figs.~\ref{acV} and \ref{y-Tc}.
When $y=5$, the resultant XRD pattern nearly conforms with those of the \ce{Sc20(C,B)8C20} and \ce{Sc2B_{1.1}C_{3.2}} phases, as shown in Fig.~\ref{xrd}, which indicates that the reaction between \ce{Sc15C19} and B with the nominal molar ratio of $1:5$ under HP resulted in these two compounds.
Moreover, given the loss-free nature of the starting materials before/after HP synthesis, we can construct the following chemical reaction:
\begin{align*}
\ce{Sc15C19} + 5\ce{B} &\longrightarrow \\ &0.55\ce{Sc20C_{2.9}B_{5.1}C20} + 2\ce{Sc2B_{1.1}C_{3.2}}
\end{align*}
wherein crystal defects such as atomic vacancies are not taken into account.
Note that the $x$ value and the molar ratio of the products are uniquely calculated, as shown in the above formula.
The obtained ratio between \ce{Sc20C_{2.9}B_{5.1}C20} and \ce{Sc2B_{1.1}C_{3.2}} $(=0.55:2)$ corresponds to a mass ratio of $1:0.4$ upon applying the molar weight of these compounds, which is roughly compatible with the XRD intensity ratio between $I_{131}^{\ce{Sc20(C,B)8C20}}$ and $I_\trm{max}^{\ce{Sc2B_{1.1}B_{3.2}}}$ in the $y=5$ sample, as depicted in Fig.~\ref{xrd}.
Furthermore, regardless of the existence ratio of \ce{Sc20C_{8-x}B_{x}C20} and \ce{Sc2B_{1.1}C_{3.2}}, the atomic ratio between Sc and C needs to be maintained throughout the HP synthesis.
This presupposition indicates that the $x$ value should be $x>2.7$ in order to coexist with the C-rich impurities, because the C/Sc ratio before the reaction is fixed at $19/15$, which supports the result of our estimation $(x \sim 5)$.
Therefore, the actual $x$ value in \ce{Sc20C_{8-x}B_{x}C20} was estimated to be about 5; this result moderately agrees with that determined as per the convex hull diagram in Fig.~\ref{convex}. 
From the optimized starting composition adopted in the arc-melting process, the content of $x$ was empirically verified to be less than 2\cite{Sc20C28-ninomiya}; however, we found that nearly half the B amount was substituted into the C layer ($8f$ site).

For the samples with $y<5$, as shown in Figs.~\ref{acV} and~\ref{MT}, we observed a systematic decrease both in $V$ and $T_\trm{c}$ values. 
Although these results suggest the formation of a solid-solution state at the $8f$ site, we previously reported that $T_\trm{c}$ is difficult to control even by tuning the nominal B content using the arc-melting method\cite{Sc20C28-ninomiya}.
From the susceptibility data in Fig.~\ref{MT}, we can confirm the broadening of the superconducting transition with decreasing $y$ values, which implies that the decrease in $T_\trm{c}$ and $V$ can be attributed to the formation of an unexpected B-deficient \ce{Sc20(C,B)8C20} structure stabilized forcibly under HP.
To clarify this issue more clearly, further structural studies based on techniques other than X-ray investigations are required.

\section{Conclusion}
By sintering \ce{Sc15B_{y}C19} $(y=1-5)$ at high temperature and pressure, we succeeded in synthesizing the \ce{Sc20C_{8-x}B_{x}C20} phase together with \ce{Sc2B_{1.1}C_{3.2}}, and we obtained the residual \ce{Sc15C19} as impurities.
Although an increase in $y$ up to $5$ led to the phase formation of \ce{Sc20(C,B)8C20} becoming predominant relative to that of \ce{Sc15C19}, the \ce{Sc20(C,B)8C20} phase suddenly disappeared at $y=7.5$, which suggests the existence of an upper $y$ limit. 
Our structural optimization obtained based on DFT calculations afforded a monotonic lattice expansion with increasing $y$, which qualitatively agreed with the experimental behavior.
To determine an optimal amount of $x$ in \ce{Sc20C_{8-x}B_{x}C20}, we assumed that the arc-melted sample contains the optimal $x$ value.
Our experiments demonstrated that the cell volume and $T_\trm{c}$ values of the $y=5$ sample are almost equal to those of the melted sample, strongly suggesting that these compounds possess an almost identical chemical composition.
We also constructed the chemical reaction scheme of the $y=5$ sample, which revealed that HP synthesis of \ce{Sc15C19 + 5B} yields the \ce{Sc20C_{8-x}B_{x}C20} $(x \sim 5)$ and \ce{Sc2B_{1.1}C_{3.2}} phases with a molar ratio of $0.55:2$.
This result is comparable with the existence ratio expected from the experimental XRD pattern.
Furthermore, our evaluation of the energy convex hull depending on $x$ revealed that the composition of \ce{Sc20C4B4C20} $(x=4)$ is the most thermodynamically stable, which is again compatible with the experimentally evaluated $x$.

\begin{acknowledgement}
The authors thank Editage (\url{http://www.editage.com}) for their English-language editing and reviewing of this manuscript.
This work was supported by a Grant-in-Aid for Scientific Research on Innovative Areas ``Quantum Liquid Crystals'' (KAKENHI Grant No. JP19H05823) from JSPS of Japan.
\end{acknowledgement}





\providecommand{\latin}[1]{#1}
\makeatletter
\providecommand{\doi}
  {\begingroup\let\do\@makeother\dospecials
  \catcode`\{=1 \catcode`\}=2 \doi@aux}
\providecommand{\doi@aux}[1]{\endgroup\texttt{#1}}
\makeatother
\providecommand*\mcitethebibliography{\thebibliography}
\csname @ifundefined\endcsname{endmcitethebibliography}
  {\let\endmcitethebibliography\endthebibliography}{}

\end{document}